\input{aipcheck}

\documentclass[
    ,final            
  ,numberedheadings 
  ]
  {aipproc}

\layoutstyle{8x11single}

\usepackage{amsmath}

\usepackage{showkeys}

\begin{document}

\newcommand\beq{\begin{equation}}
\newcommand\eeq{\end{equation}}
\newcommand\bea{\begin{eqnarray}}
\newcommand\eea{\end{eqnarray}}
\newcommand\bseq{\begin{subequations}} 
\newcommand\eseq{\end{subequations}}

\title[Classical and quantum generic \ldots]{Classical and quantum behavior \\
of the generic 
cosmological solution}


\author{Giovanni Imponente \footnote{email: gpi@physics.org}~}{
  address={Centro Studi e Ricerche ``Enrico Fermi'', 
  				Compendio Viminale  -- 00186 Roma, Italy
  				},
  altaddress={Dipartimento di Fisica, Universit\'a ``La Sapienza'', 
  Piazzale A.Moro 5, 00185 Roma, Italy}
  			}

\author{Giovanni Montani \footnote{email: montani@icra.it}~}{
  address={ENEA C.R. Frascati (U.T.S. Fusione), 
  				Via E. Fermi 45, 00044 Frascati, Roma, Italy
  			 },
  altaddress={ICRA --- International Center for Relativistic 
  Astrophysics, \\
  				Dipartimento di Fisica (G9), Universit\'a ``La Sapienza'', 
  				Piazzale A.Moro 5, 00185 Roma, Italy} 
}

\begin{abstract}
 In the present paper we generalize the original 
 work of C.W. Misner \cite{M69q} about 
the quantum dynamics of the Bianchi type IX geometry
near the cosmological singularity. 
We extend the analysis to the generic inhomogeneous 
universe by solving the super-momentum constraint
and outlining the dynamical decoupling of spatial 
points. 
Firstly, we discuss the classical evolution of the model 
in terms of the Hamilton-Jacobi approach as applied to the 
super-momentum and super-Hamiltonian constraints; 
then we 
quantize it in the approximation of a square potential well 
after an ADM reduction of the dynamics 
with respect to the super-momentum constraint only. 
Such a reduction relies on a suitable form for the 
generic three-metric tensor which allows the use 
of its three functions as the new spatial coordinates. \\
We get a functional representation of the quantum dynamics
which is equivalent to the Misner-like one when extended
point by point, since  the Hilbert space 
factorizes into $\infty^3$ independent components due 
to the  parametric role that the three-coordinates assume 
in the asymptotic potential term.\\
Finally, we discuss the conditions for having a
semiclassical behavior of the dynamics and we recognize
that this already corresponds to having mean occupation 
numbers of order $\mathcal{O}(10^2)$.

\end{abstract}

\maketitle


\section{Introduction}

The evolutionary scheme of the universe is  well 
described by the Standard Cosmological Model (SCM) 
from the light-elements nucleosynthesis 
($\sim 10^{-3} - 10^{-2}$ sec.) 
up to the present stage of the observations;
however, the very early phases of the cosmological 
dynamics (i.e. above $10^{14}-10^{15}$ GeV) requires 
more general schemes for being completely understood.\\
In fact, either the backward instability of the isotropic
universe with respect to tensor perturbations \cite{LK63, MB95}, either 
the impossibility to preserve symmetry in quantum gravity, 
strongly support the idea that the early universe must 
be represented by a generic inhomogeneous model, as discussed
in \cite{BKL70, BKL82}.

As well known, the point-like dynamics of such a generic picture
closely resembles the one characterizing the Bianchi type 
VIII and IX geometries, i.e. the so-called Mixmaster \cite{M69}.
First investigations about the classical and quantum behavior
of the Mixmaster were provided by C.W. Misner in \cite{M69} and 
\cite{M69q}, respectively; his analysis relies on 
a Hamiltonian formulation of the dynamics and on a suitable 
approximation of the spatial curvature which here plays the role 
of a potential term for the generalized coordinates
(universe volume and anisotropies).

The present work aims to generalize the main features of the 
Misner classical and quantum approaches toward the generic 
inhomogeneous universe, by virtue of the dynamical decoupling 
characterizing spatial points near the Big Bang. 

The classical inhomogeneous dynamics is investigated in 
Section \ref{gcm} via a Hamilton-Jacobi approach to the 
fundamental constraints of the theory;  the solution 
outlines  the right number of 
four physically arbitrary functions of the spatial coordinates
in the configuration space as required by the generality.

In Section \ref{adm}, before proceeding to the 
quantization,  we perform an Arnowitt-Deser-Misner (ADM)
 reduction of the super-momentum  constraints; 
in this way, we remove any dynamical role of the 
spatial gradients from the quantum behavior. 
Indeed, asymptotically to the singularity, the 
three coordinates enter the potential term  
as parameters only and then we can apply the 
 \textit{long-wavelength approximation}.
Such a framework has the physical meaning of dealing 
with inhomogeneous scales which are super-horizon 
sized; thus, when below we will speak about spatial 
points, we will refer to causal regions of the universe
which asymptotically are  dynamically decoupled. 
This fact provides a sort of causal structure to our
quantization procedure in the sense that no-interference
phenomena take place on super-horizon scales. 

In Section \ref{gqc} we give a quantum representation 
of the dynamics in terms of a functional approach which 
outlines the factorization of the Hilbert space into 
$\infty^3$ point-like components. 
The quantum dynamics relies on an adiabatic approximation
ensured by the potential term behavior, which reduces
to an infinite well; 
according to 
C.W. Misner \cite{M69q}, we model the potential 
as an infinite square box having the same measure of the 
original triangular picture. 
The volume-dependence of the wave functional acquires 
increasing amplitude and frequency of oscillations as the 
Big Bang is approached and the occupation number grows, 
respectively.

Finally, in Section \ref{semi} we discuss the conditions 
under which the WKB semiclassical limit is recognized. 
We show that this corresponds to having occupation 
numbers sufficiently greater than unity and therefore 
semiclassical packets of eigenstates would correspond
to average mean occupation number of order 
$\mathcal{O}(10^2)$.

\section{Generic Cosmological Model\label{gcm}}

When considering a cosmological solution containing 
a number of space functions such that a generic 
inhomogeneous Cauchy problem is satisfied on a 
non-singular hypersurface, we refer to it as a
generic inhomogeneous model \cite{BKL70, BKL82}. 
In the Arnowitt-Deser-Misner (ADM) formalism, 
the corresponding line element reads as
\begin{align}
ds^2 = -N(t,x)^2dt^2 + \gamma _{ij}
(dx^{i } &+ N^{i } dt)(dx^{j } +
N^{j }dt) \, ,  \\
 \gamma _{ij }&= 
 e^{q_a(t,x)} l^a_{i }l^a_{{j } }  \, , 
\label{xtx1}
\end{align}
where $N$ is the lapse-function, $N^{i }$ the 
shift-vectors and $q_a(t,x^i)$ three scalar
functions (the index $a$ is summed over $1,2,3$);
the vectors $l^a_{i }$ have components which are generic 
functions of  the spatial coordinates only, available for the 
Cauchy data. 
The general case, in which $l^a_{i }$ are 
time-dependent vectors is addressed in a following Section 
to simplify the variational principle.
It is convenient to introduce also the reciprocal 
vectors $l^i_{a }$, such that $l^a_{i }l^i_{b } =\delta^a_b$
and $l^a_{j }l^i_{a }= \delta^i_j$. \\
The dynamics of this system is summarized by the action
\begin{align}
\mathcal{S}= \int_{\Sigma^{(3)}\times\Re} d^3x dt & \left\{ p^a \partial_t q_a 
		- NH - N^{i}H_i \right\}
\label{ac} \\
H&= \frac{\chi}{\sqrt{\gamma}} \left(2 \sum_a (p^a)^2 - 
                   \sum_{m,n} p^m p^n \right) -\frac{\sqrt{\gamma}
                     }{2\chi} ~^{(3)}R  
                    \label{superh}\\
H_i &= -2\chi \partial_j \sum_m l^j_m l^m_i p^m - 
                   \chi p^j\partial_i q_j \, , 
                    \label{superm}
\end{align}
being $p^a$  the conjugate momenta to the 
variables $q_a$, $\gamma \equiv \textrm{det} \gamma_{ij}$, 
$~^{(3)}R $ the tridimensional Ricci scalar and $\chi$ the 
Einstein constant. 
By variating the action (\ref{ac}) with respect to 
the $N$ and $N^i$, we get the super-Hamiltonian and 
super-momentum constraints $H=0$ and 
$H_i=0$, respectively.\\
Let us introduce the Misner-like variables \cite{M69}
$\alpha(t,x^i)$, $\beta_+(t,x^i)$, $\beta_-(t,x^i)$
via the transformation
\bseq
\label{mlike}
\begin{align}
q_1 &= 2 \alpha + 2 \beta_+ + 2 \sqrt3 \beta_- \\
q_2 &= 2 \alpha + 2 \beta_+ - 2 \sqrt3 \beta_- \\
q_3 &= 2 \alpha -4 \beta_+  \, ;
\end{align}
\eseq
in terms of this set of configurational coordinates
the Hamiltonian constraints rewrite as
\begin{align}
H&= \chi e^{-3 \alpha} \left( - p_{\alpha}^2 + p_+^2 + p_-^2
	\right) - V \\
H_i &= - \chi \left\{ \partial_i \left(\frac{p_{\alpha}}{3} +
\frac{p_{+}}{6} + \sqrt{3}\frac{p_{-}}{6} \right) + 
\frac{1}{6} \partial_j \left[ l^j_3l^3_i 
				\left(p_+ -2\sqrt{3}p_-\right) 
			-2 \sqrt{3} ~l^j_2l^2_i p_-	\right] \right\} + \\
		&~	\qquad 	\qquad \qquad \qquad - \chi \left[ (\partial_i \alpha)p_{\alpha}
			+ (\partial_i \beta_+)p_+ + (\partial_i \beta_-)p_- \right] \\
V&=\frac{\sqrt{\gamma}}{2\chi} ~^{(3)}R    \, .
\end{align}
A detailed analysis of the potential term $V$ 
leads to \cite{LK63}
\begin{align}
\label{xtx5}
V  =   \frac{e^{\alpha}}{4 \chi} 
\left\{ a ^2_1(x^i)e^{-8\beta _+} +
a ^2_2(x^i)e^{4(\beta _+ + \sqrt{3}\beta _-)} +
a ^2_3(x^i) e^{4(\beta_+ - \sqrt{3}\beta _-)} +
W(x^i, \; \alpha, \; \beta _{\pm },\;
\partial_j \alpha ,\; \partial _j
\beta _{\pm }) \right\}  \, , 
\end{align}
where $a_i$ ($i=1,2,3$)  refer to the 
space quantities 
\begin{equation}
a_i(x^{j }) \equiv {\bf l}^i\cdot \textrm{rot} {\bf l}^i
\, ,
\label{xtx6}
\end{equation}
and we regard the operations $~\cdot ~$ and $\textrm{rot}$ 
as taken in Euclidean sense. 

To outline the relative behavior of the two terms in the  
potential as the singularity is approached for
$\alpha \rightarrow -\infty$, let us consider the 
quantities
\bseq
\begin{align}
\label{xtx7}
D & \equiv e^{3\alpha}  \\
H_1 & \equiv \frac{1}{3} + \frac{\beta _+ 
+ \sqrt{3}\beta _-}{3 \alpha}  \\ 
H_2 & \equiv \frac{1}{3} + \frac{\beta _+ 
		- \sqrt{3}\beta _-}{3\alpha}  \\
H_3 & \equiv \frac{1}{3} - \frac{2\beta _+}{3\alpha} \\
\sum _i H_i &= 1  \, . 
\end{align}
\eseq


Taking into account these definitions, the potential
$V$ rewrites as 
\begin{subequations}
\begin{align}
\label{xtx8a}
{V} = \sum _i\left( a_i^2D^{4H_i}\right) + W \\
\label{xtx8b}
W\sim \sum _{j\neq k}\mathcal{O}
\left( D^{2(H_j + H_k)} \right) 
\, .
\end{align}
\end{subequations}
Near the cosmological singularity  
$\alpha \rightarrow - \infty$ 
and $D\rightarrow 0$, so that   
the term $W$ becomes negligible.
Indeed this conclusion is supported by 
the behavior of the spatial 
gradients, which does not destroy
the feature above outlined (see \cite{K93, M95}). 

Through the canonical replacements 
\begin{align}
p_{\alpha} &=\frac{\partial S}{\partial \alpha } \\
p_{{\pm}} &=\frac{\partial S}{\partial \beta_{\pm} } \, ,
\end{align}
the classical evolution is summarized by the 
Hamilton--Jacobi system
\bseq
\begin{align}
 \label{hj1}
 &- \left( \frac{\partial S}{\partial \alpha} \right)^2
 + \left( \frac{\partial S}{\partial \beta_+} \right)^2
+ \left( \frac{\partial S}{\partial \beta_-} \right)^2   
+ V(\alpha, \beta_+, \beta_-) =0 \\
   &\frac{1}{6} \left\{ 
     \partial_i \left[ 
         2 \frac{\partial S}{\partial \alpha} 
          + \frac{\partial S}{\partial \beta_+}
          + \sqrt{3} \frac{\partial S}{\partial \beta_-} \right] 
    +\partial_j \left[ 
          l^j_3 l^3_i \left( 
            \frac{\partial S}{\partial \beta_+} 
               -2 \sqrt{3} 
           \frac{\partial S}{\partial \beta_-}\right)
          -2 \sqrt{3} l^j_2 l^2_i 
             \frac{\partial S}{\partial \beta_-}
             \right]   \right\} +
 \nonumber     \\
 \label{hj2}
  & \qquad \qquad\qquad \qquad\qquad\qquad\qquad \qquad 
   + \left\{  \left(\partial_i \alpha \right)
                 \frac{\partial S}{ \partial \alpha} +
              \left(\partial_i \beta_+ \right)     
              \frac{\partial S}{ \partial \beta_+} +
              \left(\partial_i \beta_- \right)     
              \frac{\partial S}{ \partial \beta_-} 
         \right\} =0 \, .
\end{align}
\eseq
Since sufficiently close to the cosmological 
singularity the potential term becomes negligible, 
then the solution of (\ref{hj1}) reads as
\begin{align}
\label{h1sol}
S=- \sqrt{k_+^2 + k_-^2} ~\alpha + k_+ \beta_+ 
    + k_- \beta_- \, ,
\end{align}
where $k_{\pm} = k_{\pm}(x^i)$ are arbitrary functions 
of the coordinates and the minus sign in front of the 
square root has been 
taken because we are considering an expanding Universe.

According to the Jacobi prescription, the functional 
derivatives of the above action (\ref{h1sol})
with respect to $k_{\pm}$ have to be set equal to 
stationary  quantities $c_{\pm}(x^i)$ and therefore 
we get the following expressions for $\beta_{\pm}$
in terms of $\alpha$
\begin{align}
\label{betapm}
\beta_{\pm}= \pi_{\pm} \alpha \, ,
\end{align}
where 
\bseq
\label{ppm}
\begin{align}
 & \pi_{\pm}\equiv \frac{k_{\pm}}{\sqrt{k_+^2 + k_-^2}} \\
 &\pi_+^2 + \pi_-^2 =1 \, .
\end{align}
\eseq
Substituting the solution (\ref{h1sol}) with (\ref{betapm}) 
in the Hamilton--Jacobi equation (\ref{hj2})
corresponding to the super-momentum and taking into 
account the relations (\ref{ppm}), the last sum 
cancels out leaving the equations
\begin{align}
\label{kkconstr}
 -2 \frac{k_+ + k_-}{\sqrt{k_+^2 + k_-^2}}
            \partial_i \left( k_+ + k_- \right) +
              \partial_i k_+ + \sqrt{3}\partial_i k_- 
              +
\partial_j \left[ l^j_3 l^3_i 
\left(      k_+  -2 \sqrt{3} k_-\right)
        -2 \sqrt{3} l^j_2 l^2_i k_-
              \right]     =0        \, ,
 \end{align} 
which are constraints on the spatial functions only.
The above mentioned functions $c_{\pm}(x^i)$ have been 
set equal to zero because their presence would simply correspond to 
a rescaling of the vectors $l^a_i(x^j)$. 
Thus, our solution 
contains ten arbitrary functions of the spatial coordinates, 
namely the nine components of the vectors $l^a_i$
and one of the two functions $\pi_{\pm}$. These ten 
free functions have to satisfy the three constraints 
(\ref{kkconstr}); the choice of 
the coordinate frame eliminates the arbitrariness of 
three more degrees of freedom. Therefore our solution 
is characterized by \textit{four physically} arbitrary 
functions of the spatial coordinates and, in this 
sense, it is a generic one.

Above we neglected the role of the potential 
because it influences the point-Universe evolution 
only via the bounces producing the establishment of a
new free motion (for a detailed discussion 
about the chaotic properties of the random behavior
that the point-universe performs in the potential, 
see  \cite{H94, KM97,IM01}). This effect of the potential is 
summarized by the reflection law
\beq
\sin \theta_f -\sin \theta_i =\frac{1}{2} \sin(\theta_f + \theta_i) \, ,
\eeq
where $\theta_i$ and $\theta_f$ denote the angles of incidence
and deflection, respectively,  of the point-Universe
for a bounce on one of the three equivalent 
walls of the triangular potential 
$V(\alpha, \beta_+, \beta_-)$, see Figure \ref{fig:domain}.

The results of this Section show how the generic cosmological 
solution toward the Big-Bang is isomorphic, point by point
in space, to the one of the Bianchi types VIII and IX models
\cite{B82} because the spatial coordinates are involved 
in the problem only as parameters.

\section{ADM reduction of the dynamical problem\label{adm}}

In the last Section we have discussed a generic cosmological 
model in which the vectors $l^a_{i}$ appearing in (\ref{xtx1}) were 
functions of the coordinates $x^{j}$ only. 
It is interesting to discuss a more generic framework, 
that is to allow the more general dependence
$ l^a_{i} = l^a_{i} (t, x^{j})$. \\
To this purpose, let us rewrite such vectors in the form \cite{BM04}
\beq\label{ldiy}
l^a_{i} =  O^a_b \partial_{i} y^b \, ,
\eeq
where $O^a_b=O^a_b(x^k)$ is a $SO(3)$ matrix and 
$y^b=y^b(t,x^k)$ are three scalar functions.
Considering  the metric tensor (\ref{xtx1}) rewritten as 
\beq
\gamma_{ij}=e^{q_a}\delta_{ad}
O^a_b O^d_c \partial_i y^b \partial_j y^c \, , 
\eeq
with the indexes $(a,b,c,d,i,j)$ running 
as $1,2,3$, the action (\ref{ac})becomes
\begin{equation}
\label{acstand}
	\mathcal{S}=\int_{\Sigma^{(3)}\times\Re}
	d^3 x dt \left(p_a\partial_t q^a+\Pi_d\partial_t y^d 
	- NH - N^i H_i\right)\,,
\end{equation}

\begin{figure}
  \includegraphics[height=.3\textheight]{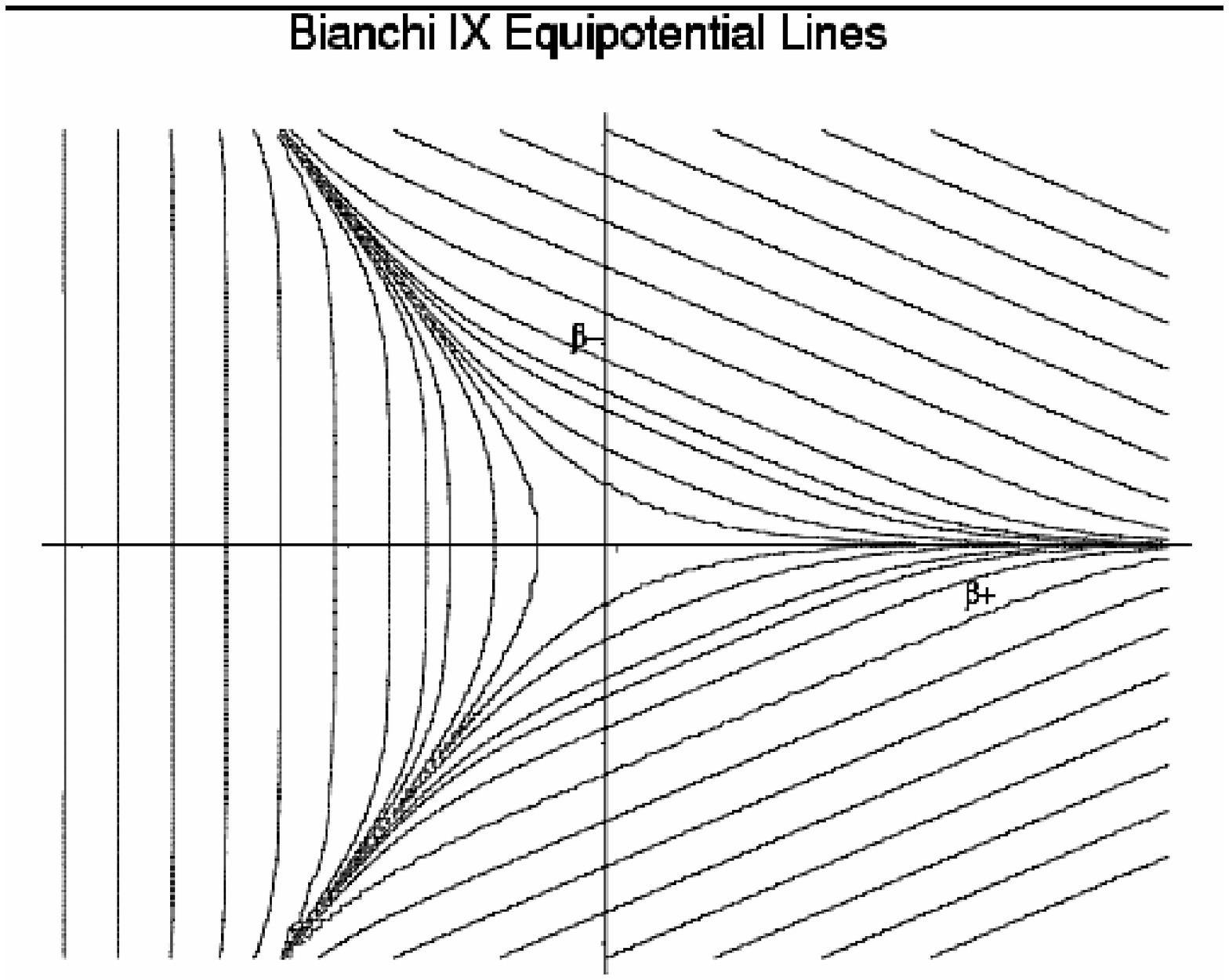}
\label{fig:domain}
 \caption{Equipotential lines showing the potential 
 domain $\Gamma_H$, dynamically 
 closed in the corners: this visualization
 is in the $\beta_+$, $\beta_-$ plane of the Misner variables.
 Asymptotically, the walls flatten to a triangular well, as in 
 Equation (\ref{vu}).}
\end{figure}

where, apart some details of the potential coordinate dependence,
$H$ has the same expression as (\ref{superh}), 
while the super-momentum explicitly reads
	\begin{equation}	
	\label{ssuperm} 
	 H_i=\Pi_c \partial_i y^c +p_a \partial_i q^a  
	  +2p_a\left(O^{-1}\right)^b_a\partial_i O^a_b \, ,
\end{equation}
$\Pi_d$ being the conjugate momenta of the 
variables $y^d$. \\
Summarizing, we have ten independent functions characterizing
the dynamical system: the 3 degrees of freedom 
of the coordinate choice $y^a$, 3 scale factors
$q^a$, 3 components of the shift vector $N^i$ and
the lapse function $N$. \\
The variation of the action (\ref{acstand}) with respect 
to $N$ and $N^i$ leads to the usual  constraints
$H=0$ and $H_i=0$, respectively. 
Choosing  $y^a$ as spatial coordinates, 
i.e. $y^a=y^a(t,x)$ and $\eta=t$, 
we can diagonalize and solve the 
constraint (\ref{ssuperm}) as follows
\begin{equation}
\label{sol_supm}
	\Pi_b=-p_a\frac{\partial q^a}{\partial y^b}-2p_a
	\left(O^{-1}\right)^c_a\frac{\partial O^a_c}{ \partial y^b}	\, .
\end{equation}
Given  $|J|$ as the Jacobian of the change of coordinates,
the  relations 
\begin{equation}
	\begin{cases}
	q^a(t,x)\to {q^{\prime}}^a(\eta,y)\cr
	p_a(t,x)\to {p^{\prime}}_a(\eta,y)={\displaystyle \frac{p_a(\eta,y)}{|J|}} \cr
	\displaystyle\frac{\partial}{ \partial t}\rightarrow 
	\frac{\partial y^b}{\partial t}
	\frac{\partial}{ \partial y^b}+
	\frac{\partial}{ \partial \eta}\cr
	\displaystyle\frac{\partial}{ \partial x^\alpha}\rightarrow 
	\frac{\partial y^b}{ \partial x^\alpha}
	\frac{\partial}{ \partial y^b}\cr
	\end{cases}
\end{equation}
allow to write the (reduced) action as 
\begin{equation}
	\label{act_red1}
	\mathcal{S}_{RED}=\int_{\Sigma^{(3)}\times\Re}d\eta d^3 y 
	\left(p_a\partial_\eta q^a - 2p_a\left(O^{-1}\right)^c_a
	\partial_\eta O^a_c-NH\right)	\,.
\end{equation} 

When approaching the cosmological singularity, 
the potential term (\ref{xtx8a}) can be expressed as
\begin{equation}
\label{vu}
	V=\sum_a{\Theta(H_a)}\, ,
\end{equation}
being
\begin{equation}
	\label{ftheta}
	\Theta(x)=\begin{cases}
	+\infty\ &  $if$\  x>0,\cr 0\ &  $if$\  x<0 \, .\cr
	\end{cases}
\end{equation}
The dynamics of the universe is decoupled for 
each space point and the point-universe
moves in the domain $\Gamma_H$ as outlined by the 
triangular equipotential lines of Figure \ref{fig:domain}, 
which is dynamically closed also in the corners.

The vanishing behavior of the potential inside the domain of 
definition implies that, near the singularity, 
$\partial p_a/\partial \eta=0$. 
Therefore the term 
$p_a\left(O^{-1}\right)^c_a	\partial_\eta O^a_c$
in (\ref{act_red1}) reads as an exact time differential term, 
and unaffects the asymptotic dynamics  whose 
variational principle takes the form \cite{BM04}
\beq
\label{s_adm}
\mathcal{S}^{\prime}_{RED}=\int_{\Sigma^{(3)}\times\Re}d\eta d^3 y 
\left(p_a\partial_\eta q^a -NH\right) \, ,
\eeq
and the dynamics is described by $H$ only.\\

\section{Generic Quantum Cosmology\label{gqc}}

Let us now analyze the quantum dynamics corresponding 
to a generic cosmological model within a canonical framework, 
by replacing the canonical variables with the corresponding 
operators and implementing the  Hamiltonian constraints
on the state functional describing the system, 
$\Psi=\Psi(\alpha, \beta_+, \beta_-)$.
Therefore we adopt the representation
\begin{align}
p_{\alpha} \rightarrow - i \hbar \frac{\delta ~}{\delta \alpha} \, ,
\qquad  
p_{\beta_{\pm}} \rightarrow - i \hbar \frac{\delta ~~}{\delta \beta_{\pm}} \, ,
\end{align}
and implementing $H$ to an operator the quantum dynamics 
is described by the Wheeler-deWitt equation

\beq
\label{wdw}
\hat{H}\psi= \chi e^{-3 \alpha}\left(-\hbar^2\right)
\left[ - \frac{\delta^2}{\delta \alpha^2}
	+\frac{\delta^2}{\delta \beta_+^2}
	+\frac{\delta^2}{\delta \beta_-^2}	  \right]\psi
-e^{\alpha}V\psi =0 \, . 
\eeq
As discussed at the end of the preceding 
Section, since the  asymptotically 
the potential term remains isomorphic
to the Bianchi IX one, point by point in space, we can 
find a solution in the form 
\beq
\psi= \sum_n \Gamma_n(\alpha)\varphi_n(\alpha, \beta_+, \beta_-) \, ,
\eeq
where the coefficients $\Gamma_n$ 
\beq
\label{coeff_c}
\Gamma_n = \textrm{exp} 
\left( \int d^3x \ln C_{n(x^i)}(\alpha)  \right) =
\prod_{x^i} C_{n(x^i)}(\alpha)
\eeq
keep trace of the inhomogeneities  of the system and the 
index $n$ has to be regarded as a space function $n(x^i)$. \\
Thus, equation (\ref{wdw}) is reduced to the ADM eigenvalue problem
\beq
\label{eeii}
\left[-\frac{\delta^2}{\delta \beta_+^2}
	-\frac{\delta^2}{\delta \beta_-^2}	+
e^{4\alpha}V  \right]\phi
 = E_n^2(\alpha)\phi  \,.
\eeq
According to \cite{M69q}
we approximate the triangular infinite walls of the 
potential by a box having the same measure 
and then we find  the eigenvalues $E_n$
\beq
\label{eig_1}
E_n(\alpha)= \pi \left(\frac{4}{3^{3/2}}\right)^{1/2} 
\frac{\mid n \mid}{\alpha} \, ,
\eeq
where $n^2\equiv n_+^2+ n_-^2$, being 
$n_+,n_- \in \mathcal{N}$ the two independent quantum 
numbers corresponding to the  variables $\beta_+, \beta_-$, 
respectively.

\begin{figure}
  \includegraphics[height=.3\textheight]{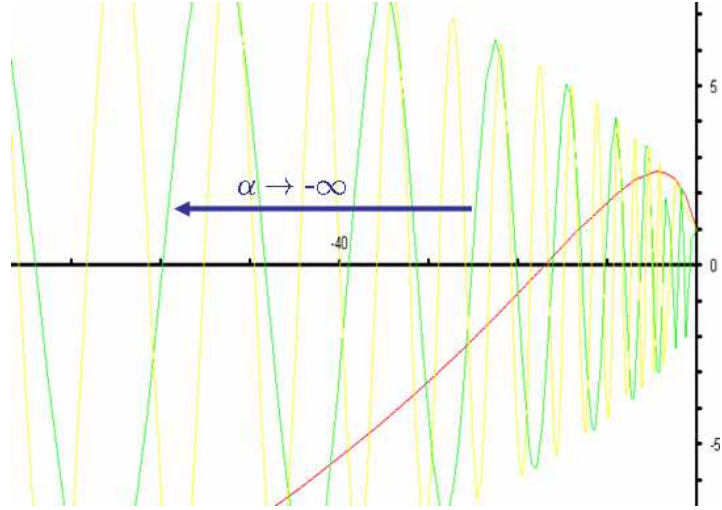}
\label{fig:c_alpha}
 \caption{Behavior of the solution $C_n(\alpha)$ for three
 different values of the parameter $k_n=1, 15,30$. 
 The bigger $k_n$, the higher the frequency of oscillation. }
\end{figure}

Substituting the expression for 
$\Gamma_n$ in equation (\ref{wdw}) 
we get the differential equation 
\beq
\sum_n(\partial^2_{\alpha} C_n) \varphi_n + 
\sum_n C_n (\partial^2_{\alpha} \varphi_n)  +
2\sum_n (\partial_{\alpha} C_n )(\partial_{\alpha}\varphi_n)+
\sum_n E_n^2 C_n \varphi_n =0 \, , 
\eeq
which, in the limit of the Misner adiabatic approximation 
of neglecting $\partial_{\alpha}\phi_n$ 
(i.e. $\phi \sim \phi(\beta_+, \beta_-)$), simplifies to
%
\beq
\label{e_diff2}
\hbar^2\frac{d^2C_n}{d\alpha^2} +\frac{k^2}{\alpha^2}C_n=0
\eeq
where 
\beq
k_n^2=\left(\frac{2}{3} \pi\right)^{3/2}\mid n\mid^2 \hbar^2 \, .
\eeq
The above equation is solved by $C_n(\alpha)$ in the form
\beq
C_n(\alpha) =C_1\sqrt{\alpha} \sin\left(\frac{1}{2} \sqrt{p_n} 
\ln \alpha\right)+
C_2\sqrt{\alpha} \cos\left(\frac{1}{2} \sqrt{p_n} \ln \alpha\right)\, , 
\eeq
where $\sqrt{p_n}=\sqrt{k_n^2-1}$. Figure \ref{fig:c_alpha} shows
the behavior of $C_n(\alpha)$ for various values of the 
parameter $k_n$. 
Such wave function  at fixed $x^i$
behaves like an oscillating profile whose frequency 
increases with occupation number $n$ and approaching the Big Bang,
while the amplitude  depends on the $\alpha$ variable only.

\section{Semiclassical limit of the quantum dynamics\label{semi}}

It is now interesting to consider the evolution of the system 
in the semiclassical limit, that is to investigate the 
dynamical behavior in the vicinity of the Big Bang
within the WKB approximation of the dynamics
(for a discussion of the semiclassical limit in the homogeneous 
case, see \cite{IM03a}). 
To this purpose, let us search for a solution of the 
wave equation (\ref{e_diff2}) in the form 
\beq
\label{c_semicl}
C(\alpha)=\sqrt{\rho} e^{i \sigma/\hbar} \, ,
\eeq
which  reduces such equation to the system of the
real and complex parts  as
\begin{subequations}
\begin{align}
\label{semi_r}
  - \left( \frac{d\sigma}{d\alpha} \right) ^2
+ \frac{k^2}{\alpha^2}  
+
\hbar^2 &\left[ -
 \left( \frac{1}{2 \rho}\frac{d\rho}{d\alpha}  \right)^2 
  + \frac{1}{2 {\rho}}  \frac{d^2\rho}{d\alpha^2}  \right]
=0   \\
\label{semi_i}
\hbar &  \frac{d}{d\alpha} 
\left({\rho} \frac{d\sigma}{d\alpha} \right)=0 \, ,
\end{align}
\end{subequations}
respectively. 
In the semiclassical limit, neglecting the 
term in square brackets in Equation (\ref{semi_r}),
we obtain the solution 
\begin{subequations}
\begin{align}
\sigma(\alpha) &= k\ln \alpha \, , \\
\rho(\alpha)&= const. \times \alpha \, .
\end{align}
\end{subequations}
This behavior remains valid as far as the term 
in $\hbar^2$ is negligible and therefore 
we get
the constraint   
\begin{equation}
k^2 \gg \frac{\hbar^2}{4} \, ,
\end{equation}
i.e. $n \gg 1$. 
As soon as this constraint is fulfilled, 
the solution remains semiclassical close to the Big Bang
and we see that, in the sense of the WKB approximation, 
this would imply packets of eigenfunctions whose 
mean quantum number $<n>$ can be of order 
$\mathcal{O}(10^2)$ only.

\section{Concluding remarks}

By our analysis, we gave an inhomogeneous extension
of the classical and quantum Mixmaster dynamics. 
Following the original point of view of C.W. Misner, 
we achieved a quantum description of the model 
by approximating the potential term as an infinite 
square box. 

The main issue of our investigation has shown 
how the removal of the spatial gradients dynamics 
is possible in the asymptotic regime. 
This result comes out from an ADM reduction of the 
super-momentum constraints and from the parametric
role that the three-coordinates assume in the potential term. 

On the quantum level, this achievement turned out into the 
factorized structure of the Hilbert space; from a physical 
point of view, this implied to deal with independent 
quantum behaviors within each cosmological horizon.

Finally, we discussed the conditions allowing a WKB 
semiclassical approach to the dynamics and we fixed 
as condition to have occupation numbers much greater 
than unity.


\begin{theacknowledgments}
  One of us, G. Imponente, would like to thank the 
  Royal Astronomical Society (RAS), UK, and the 
  Institute of Physics (IoP), UK, 
  for having partially supported  this work.
  
\end{theacknowledgments}





\begin{thebibliography}{9}


\bibitem{M69q} C.W. Misner,
\textit{Phys. Rev.} \textbf{186}, 1319-1327 (1969)

\bibitem{LK63}  E.M. Lifshitz and I.M. Khalatnikov, 
\textit{Adv. Phys.}  
\textbf{12},  185 (1963).



\bibitem{MB95} C.P. Ma and E. Bertschinger, 
\textit{Astrophys.J.},
 \textbf{455}, 7 (1995).



\bibitem{BKL70} V.A. Belinski, I.M. Khalatnikov, and E.M. Lifshitz, 
\textit{Adv. Phys.} \textbf{ 19}, 525 (1970).


\bibitem{BKL82} V.A. Belinski, I.M. Khalatnikov, and E.M. Lifshitz, 
\textit{Adv. Phys.} \textbf{31}, 639 (1982).





\bibitem{M69} C.W. Misner, 
\textit{Phys. Rev. D}  \textbf{186}, 1319 (1969).



\bibitem{K93}
A.A. Kirillov,
\textit{Zh.\ Eksp.\ Theor.\ Fiz.},\textbf{103}, 721-729, (1993).


\bibitem{M95}
G. Montani, \textit{Class. and Quantum Grav.}, 
\textbf{12}, 2505, (1995).

\bibitem{H94}
\textit{Deterministic Chaos in General Relativity}, 
edited by D.Hobill, A. Burd, and A. Coley, 
World Scientific, Singapore, (1994).



\bibitem{KM97}
	A. A. Kirillov and G. Montani, 
\textit{JETP Letters} \textbf{66}, 475 (1997).



\bibitem{IM01} G. Imponente and G. Montani
\textit{Phys. Rev. D} \textbf{63}, 103501 (2001).





\bibitem{B82} 
J.D. Barrow
\textit{Phys. Rep.} {\bf 85}, 1 (1982). 






\bibitem{BM04} R. Benini and G. Montani, 
\textit{Phys. Rev. D} \textbf{70}, 103527 (2004). 



\bibitem{IM03a}
G. Imponente and G. Montani
{\it Int. Journ. Mod. Phys. D}, 
\textbf{12}, 977  (2003).


















\end{thebibliography}



\end{document}